\documentstyle [fleqn,12pt] {article}
\def\mscript#1{\mbox{\scriptsize$#1$}}

\begin{document}
\pagestyle{empty}

\vspace*{2cm}
\begin{center}
{\bf Global Phase Structure of the Heterotic 
Thermal String} \\
\vspace{1cm} 
H. Fujisaki, S. Sano and K.Nakagawa$^{*}$

\vspace{5mm}

{\it Department of Physics, Rikkyo University, Tokyo 171}\\
$^{*}$ {\it Faculty of Pharmaceutical Sciences, Hoshi University, Tokyo 142}\\
\vspace*{3cm}
{\bf ABSTRACT}
\end{center}

\indent The global phase structure of the $D = 10$ 
heterotic thermal string ensemble is studied through the infrared 
behaviour of the one-loop dual symmetric cosmological constant in 
proper reference to the thermal duality symmetry as well as the 
thermal stability of modular invariance within the general framework 
of the thermofield dynamics.    
\newpage
%%%%%%%%
%%%%%%%%
\pagestyle{plain}
\setcounter{page}{1}
\indent Elaboration of thermal string theories based upon the thermofield 
dynamics (TFD) \cite{umezawa} has gradually turned out to be a good 
practical subject of high energy physics \cite{leblanc1} - 
\cite{nakagawa}.  In a previous paper of ourselves \cite{fujisaki4}, we 
have succeeded in shedding some light upon the global phase structure of 
the thermal string excitation in proper reference to the thermal duality 
relation \cite{obrien} -- \cite{osorio} for the $D = 26$ closed bosonic 
thermal string theory within the TFD framework.  In the present 
communication, the global phase structure of the $D = 10$ heterotic thermal string 
ensemble is examined $\grave{a}\: la$ O'Brien and Tan 
\cite{obrien} on the basis of the TFD algorithm through the infrared 
behaviour of the one-loop cosmological constant in proper reference to the thermal 
duality symmetry as well as the thermal stability of modular invariance.

Let us start with describing the one-loop cosmological constant 
$\Lambda(\beta)$ at any finite temperature $\beta^{-1} = kT$ as
\vspace{5mm}
\begin{equation}
\Lambda (\beta) = \frac{\alpha^\prime}{2} \lim_{\mu^2 \rightarrow 0} {\rm 
Tr} 
\left[ \int_{\infty}^{\mu^2} dm^2 \left( \Delta^\beta_B (p, P; m^2) + 
\Delta^{\beta}_{F} (p, P; m^2) \right) \right]
\end{equation}

\vspace{5mm}
\noindent for the $D = 10$ heterotic thermal string theory in the TFD 
framework, 
where $\alpha^\prime$ means the slope parameter, $p^\mu$ 
reads loop momentum and $P^I$ lie on the even self-dual root lattice $L = 
\Gamma_8 \times \Gamma_8$ for the exceptional group $G = E_8 \times 
E_8$ \cite{green}.  Here the thermal propagator 
$\Delta^\beta_{B[F]} (p, P; m^2)$ of the free closed bosonic [fermionic] 
string is expressed at $D = 10$ as
 
\begin{eqnarray}
\lefteqn{\Delta^\beta_{B[F]}(p, P; m^2) = \int_{-\pi}^{\pi} 
\frac{d\phi}{4\pi} \; {\rm e}^{i\phi \left( N - \alpha - \bar{N} + 
\bar{\alpha} -1/2 \cdot \sum_{I=1}^{16} (P^I)^2 \right) }}
\nonumber \\
& & \times \Biggl( \left[ \raisebox{-1ex}{$\stackrel {\textstyle 
+}{\mscript{[}-\mscript{]}}$} \int_{0}^{1} dx + \frac{1}{2} \sum_{n=0}^{\infty} \frac{\delta [\alpha^\prime 
/2 \cdot p^2 + \alpha^\prime /2 \cdot m^2 + 2(n - \alpha)]}{{\rm e}^{\beta 
|p_0|} \raisebox{-1ex}{$\stackrel{\textstyle -}{\mscript{[}+\mscript{]}}$} \; 1} 
\oint_{c} dx \right] \nonumber \\
& & \times x^{\alpha^\prime /2 \cdot p^2 + N - \alpha + \bar{N} - 
\bar{\alpha} + 1/2 \cdot \sum_{I=1}^{16} (P^I)^2 + \alpha^\prime /2 \cdot 
m^2 - 1} \Biggr) \quad ,
\end{eqnarray}

\vspace{5mm}
\noindent where $N$ [$\bar{N}$] denotes the number operator of the right- 
[left-] 
moving mode, the intercept parameter $\alpha$ [$\bar{\alpha}$] of the 
right- [left-] mover is eventually fixed at $\alpha = 0$  
$[\bar{\alpha} = 1]$ and the contour $c$ is taken as the unit circle 
around the origin.  We are then eventually led to 
the modular parameter integral representation of $\Lambda (\beta)$ at $D = 10$ 
as follows \cite{obrien}:
\vspace{5mm}
\begin{eqnarray}
\Lambda (\beta) & = & -8(2\pi \alpha^\prime)^{-D/2} \int_{E} 
\frac{d^2\tau}{2\pi \tau_2^2} \; (2\pi \tau_2)^{-(D-2)/2}\; {\rm 
e}^{2\pi i \bar{\tau}} \left[ 1 + 480 \sum_{m=1}^{\infty} 
\sigma_7(m)\bar{z}^m \right] \nonumber \\
& & \times \prod_{n=1}^{\infty} (1 - \bar{z}^n)^{-D-14} \left( 
\frac{1 + z^n}{1 - z^n} \right)^{D-2} \sum_{\ell \in Z; {\rm odd}} \exp \left[ 
- \frac{\beta^2}{4\pi \alpha^\prime \tau_2}\; \ell^2 \right] 
\rule{0cm}{1cm} \quad ,
\label{eq:Lambda2} 
\end{eqnarray}

\vspace{5mm}
\noindent where $\stackrel{[\normalsize{-}]}{\tau} = \tau_1 
\raisebox{-1ex}{$\stackrel{\normalsize{+}}{\mscript{[}\!-\!\mscript{]}}$} 
i\tau_2$, 
$z = x {\rm e}^{i\phi} = {\rm 
e}^{2\pi i \tau}$, $\bar{z} = x{\rm e}^{-i \phi} = {\rm e}^{-2\pi i 
\bar{\tau}}$, $\alpha \; [\bar{\alpha}]$ has been fixed at $\alpha = 
0 \; [\bar{\alpha} = 1]$, $E$ means the half-strip integration 
region in the complex $\tau$ plane, i.e. $-1/2 \leq \tau_1 \leq 1/2$; 
$\tau_2 > 0$, and the full use has been made of an explicit expression of 
the theta 
function $\Theta_{\Gamma_8 \times \Gamma_8}$ of the root lattice 
$\Gamma_8 \times \Gamma_8$ \cite{green}.  Accordingly, the $D = 10$ 
thermal amplitude $\beta \Lambda (\beta)$ is identical in every detail 
with the ``$E$-type'' representation of the thermo-partition function 
$\Omega_h (\beta)$ of the heterotic string in ref. \cite{obrien} as 
required from the equivalence of the thermal cosmological constant and 
the free energy amplitude.  As can readily be envisaged from eq. 
($\!$~\ref{eq:Lambda2}), the ``$E$-type'' thermal amplitude $\Lambda 
(\beta)$ 
is not modular invariant and annoyed with ultraviolet 
divergences at the endpoint $\tau_2 \sim 0$ for $\beta \leq 
\beta_H = (2 + \sqrt{2}) \pi \sqrt{\alpha^\prime}$, where $\beta_H$ 
reads the inverse Hagedorn temperature of the heterotic thermal 
string.  It is parenthetically mentioned that the thermal amplitude 
$\Lambda (\beta)$ is infrared convergent at $\tau_2 \rightarrow 
\infty$ for any value of $\beta$.  For further details of the 
asymptotic behaviour of $\Lambda (\beta)$, we merely refer to ref. 
\cite{obrien}.    

Our prime concern is reduced to regularizing the thermal amplitude 
$\Lambda (\beta)$ {\it $\grave{a}$ la} ref. \cite{obrien} as well as 
ref. \cite{osorio} through transforming the physical information in 
the ultraviolet region of the half-strip $E$ into the ``new-fashioned'' 
modular invariant amplitude.  Let us postulate the one-loop dual symmetric thermal cosmological 
constant $\bar{\Lambda} (\beta; D)$ at any space-time dimension $D$  as an integral over the 
fundamental domain $F$, i.e. $-1/2 
\leq \tau_1 \leq 1/2 \: ; \tau_2 > 0 \: ; |\tau| > 1$, of the modular 
group 
$SL(2, Z)$ as follows \cite{obrien} :  
\begin{equation}
\bar{\Lambda} (\beta; D) = - 
\frac{16}{\beta} (2\pi \alpha^\prime)^{-D/2} \sum_{(\sigma, \rho)} 
\int_{F} 
\frac{d^2\tau}{2\pi \tau_2^2}\; B(\bar{\tau}, \tau; 
D) A_{\sigma \rho} (\tau; D) D_{\sigma \rho} (\bar{\tau}, \tau; \beta) \quad,
\end{equation}
where
\vspace{5mm}
\begin{eqnarray}
B(\bar{\tau}, \tau; D) & = & - \frac{1}{8} (2 \pi \tau_2)^{-(D-2)/2}\;  
\bar{z}^{-(D+14)/24} z^{-(D-2)/24} \nonumber \\
& & \times \left[ 1 + 480 \sum_{m=1}^{\infty} \sigma_7 (m) 
\bar{z}^m \right] \prod_{n=1}^{\infty} (1 - \bar{z}^n)^{-D-14} 
(1 - z^n)^{-D+2}, 
\end{eqnarray}

\begin{eqnarray}
\left( \begin{array}{c}
A_{+-}(\tau; D) \rule[-2mm]{0mm}{8mm} \\ A_{-+}(\tau; D) \rule[-2mm]{0mm}{8mm} \\ A_{--}(\tau; D) 
\end{array} \right) 
= 8 \left( \frac{\pi}{4} \right) ^{(D-2)/6} 
\left( \begin{array}{l}
-[\theta_2(0, \tau)/\theta_{1}^{\prime}(0, \tau)^{1/3}]^{(D-2)/2}
\rule[-2mm]{0mm}{8mm} \\
-[\theta_4(0, \tau)/\theta_{1}^{\prime}(0, \tau)^{1/3}]^{(D-2)/2}
\rule[-2mm]{0mm}{8mm} \\
\; [\theta_3(0, \tau)/\theta_{1}^{\prime}(0, \tau)^{1/3}]^{(D-2)/2}
\rule[-2mm]{0mm}{8mm}
\end{array} \right) \quad ,
\end{eqnarray}

\vspace{5mm}
\begin{equation}
D_{\sigma \rho}(\bar{\tau}, \tau; \beta) = 
C_\sigma^{(+)}(\bar{\tau}, \tau; \beta) 
+ \rho C_\sigma^{(-)}(\bar{\tau}, \tau; \beta) \quad,
\end{equation}

\vspace{5mm}
\begin{equation}
C_\sigma^{(\gamma)}(\bar{\tau}, \tau; \beta) = (4\pi^2\alpha^\prime 
\tau_2)^{1/2} \sum_{(p, q)} \exp \left[ - \frac{\pi}{2} \left( 
\frac{\beta^2}{2\pi^2\alpha^\prime} p^2 + \frac{2\pi^2\alpha^\prime}
{\beta^2} q^2 \right) \tau_2 + i\pi pq \tau_1 \right] ,
\end{equation}

\vspace{5mm}
\noindent the signatures $\sigma, \rho$ and $\gamma$ read $\sigma, \rho = +, 
- ; \; -, + ; \; -, -$ and $\gamma = +, -$, respectively, the summation 
over $p \; [q]$ is restricted by $(-1)^p = \sigma \; [(-1)^q = \gamma]$ 
and the 
explicit use has been made of the Jacobi theta functions $\theta_j(0, 
\tau);\; j = 1, 2, 3, 4$ as well as the Poisson resummation formula.  It 
is 
almost needless to mention that the $D = 10$ thermal amplitude 
$\beta\bar{\Lambda}(\beta; D = 10)$ is literally reduced to the 
``$D$-type'' representation of the thermo-partition function 
$\Omega_h(\beta)$ in ref. \cite{obrien} which in turn guarantees 
$\bar{\Lambda}(\beta; D = 10) = \Lambda (\beta)$ as expected from 
self-consistency.  Let us now examine 
the algebraic structure of the ``$D$-type'' thermal amplitude 
$\bar{\Lambda}(\beta; D)$ with the arithmetic aid of Appendix B in 
ref. \cite{obrien}.  Typical theoretical observations are as follows:  The 
scalar product $\sum_{(\sigma, \rho)} A_{\sigma \rho} D_{\sigma \rho}$ is 
invariant under permutations of the signature, irrespective of the values 
of $\beta$ and $D$, not only for the shifting transformation $\tau 
\rightarrow \tau + 1$ but also for the inversion $\tau \rightarrow 
-\tau^{-1}$.  In addition, $B(\bar{\tau}, \tau; D)$ is invariant, 
irrespective of the value of $D$, under the action of any modular 
transformation.  Accordingly, the thermal amplitude 
$\bar{\Lambda}(\beta; D)$ is manifestly modular invariant and 
free of ultraviolet divergences for any value of $\beta$ and $D$.  As a 
matter of fact, moreover, the generalized duality symmetry \cite{obrien} 
$C_\sigma^{(\gamma)}(\bar{\tau}, \tau; \beta) = 
C_\gamma^{(\sigma)}(\bar{\tau}, \tau; \tilde{\beta})$ holds for any value 
of $D$, where $\tilde{\beta} = 2\pi^2 \alpha^\prime /\beta$.  If and only 
if $D = 10$, on the other hand, the scalar 
product $\sum_{(\sigma, \rho)} A_{\sigma \rho} D_{\sigma \rho}$ is 
invariant under the thermal duality transformation $\beta \leftrightarrow 
\tilde{\beta}$ as a simple and natural consequence of the Jacobi identity 
$\theta_2^4 - \theta_3^4 + \theta_4^4 = 0$ for the theta functions.  
If and only if $D =  10$, then, the thermal duality relation $\beta 
\bar{\Lambda}(\beta; D) = 
\tilde{\beta}\bar{\Lambda}(\tilde{\beta}; D)$ is manifestly satisfied 
for the thermal amplitude $\bar{\Lambda}(\beta; D)$, irrespective of 
the value of $\beta$.  

Let us recall to our remembrance that $\theta_{1}^{\prime -1/3} \sim {\rm 
e}^{\pi \tau_2/12};\; \theta_2 \sim 0;\; \theta_3 \sim 1;\; \theta_4 \sim 
1$ 
near $\tau_2 \rightarrow \infty$.  The infrared behaviour of the 
thermal cosmological constant $\bar{\Lambda}(\beta; D)$ 
is then asymptotically described at $\tau_2 \rightarrow \infty$ as  
\begin{eqnarray}
\bar{\Lambda}(\beta; D) & = & - 64 \sqrt{2}\: (8\pi^2 \alpha^\prime)^{-D/2} 
\sum_{(p, q)} \int_{- \frac{1}{2}}^{\frac{1}{2}} d\tau_1 \exp [i\pi 
pq\tau_1] \, \sqrt{\frac{\tilde{\beta}}{\beta}} 
\int_{\sqrt{1-\tau_1^2}}^{\infty} d\tau_2 \; \tau_2^{-(D+1)/2} \nonumber \\
& & \times \exp \left[ - \frac{\pi}{2} \; \tau_2 \left( 
\frac{\beta}{\tilde{\beta}} \; p^2 + \frac{\tilde{\beta}}{\beta} \; q^2 - 
\frac{5}{12} (D - 10) - 6 \right) \right] \quad ,
\label{eq:Lambda3}
\end{eqnarray}

\vspace{5mm}
\noindent where $p, q = \pm 1; \pm 3; \pm 5; \cdots$.  As can easily be seen 
from 
eq. ($\!$~\ref{eq:Lambda3}), uniform convergence of the thermal
 amplitude $\bar{\Lambda}(\beta; D)$ is assured at any value of $\beta$ if and 
only if $D < 2/5$.  Of principal concern with us is the case $D = 10$, 
anyhow.  Infrared convergence of the $D = 10$ TFD amplitude 
$\bar{\Lambda}(\beta; D = 10)$ is then guaranteed if and only if 
either $(2 + \sqrt{2})\pi \sqrt{\alpha^\prime} = \beta_H < \beta < 
\infty$ or $0 < \beta < \tilde{\beta}_H = (2 - \sqrt{2}) \pi 
\sqrt{\alpha^{\prime}}$, where 
$\tilde{\beta}_H$ reads the inverse dual Hagedorn temperature of the 
heterotic thermal string.   
Explicit calculation of the $\tau_2$ integral 
in eq. ($\!$~\ref{eq:Lambda3}) is readily performed for the case $D < 
2/5$.  The resultant TFD amplitude indeed obeys the thermal 
duality symmetry $\beta \bar{\Lambda}(\beta; D) = 
\tilde{\beta}\bar{\Lambda}(\tilde{\beta}; D)$ and brings forth the 
correct analytic continuation from $D < 2/5$ to higher values of $D$, 
i.e. $D = 10$.  We can therefore define the dimensionally regularized, 
$D = 10$ one-loop dual symmetric thermal cosmological constant 
$\hat{\Lambda}(\beta)$ by 
\begin{eqnarray}
\hat{\Lambda}(\beta) & = &  -\frac{2}{\beta} (8\pi 
\alpha^\prime)^{-(D-1)/2} \sum_{(p, q)} 
\int_{- \frac{1}{2}}^{\frac{1}{2}} d\tau_1 \exp[i\pi pq \tau_1] \nonumber 
\\
& & \times \left( \frac{\beta^2}{2\pi^2\alpha^\prime} \: p^2 + 
\frac{2\pi^2\alpha^\prime}
{\beta^2} \: q^2 - 6 \right) ^{(D-1)/2} \nonumber \\
& & \times \Gamma \left[ - \frac{D - 1}{2}\: ,\; \frac{\pi}{2} \sqrt{1 - 
\tau_1^2} \left( \frac{\beta^2}{2\pi^2 \alpha^\prime} \: p^2 + 
\frac{2\pi^2 
\alpha^\prime}{\beta^2} \: q^2 - 6 \right) \right]\; ;\quad D = 10 ,
\label{eq:Lambda5}
\end{eqnarray}

\vspace{5mm}
\noindent irrespective of the value of $\beta$, where $\Gamma$ is the 
incomplete gamma function of the second kind.  The dimensionally 
regularized, thermal cosmological constant $\hat{\Lambda} (\beta)$
 manifestly satisfies the thermal duality relation $\beta 
\hat{\Lambda}(\beta) = \tilde{\beta} \hat{\Lambda}(\tilde{\beta})$ in 
full accordance with the thermal stability of the 
fundamental properties such as modular invariance.  

Let us examine the singularity structure of the dimensionally 
regularized, $D = 10$ dual symmetric thermal amplitude 
$\hat{\Lambda}(\beta)$.  The position of the singularity $\beta_{|p|, 
|q|}$ is determined by solving $\beta/\tilde{\beta} \cdot p^2 + 
\tilde{\beta}/\beta \cdot q^2 - 6 = 0$ for every allowed $(p, q)$ in eq. 
($\!$~\ref{eq:Lambda5}).  We then obtain two sets of solutions 
with $|pq| \leq 3$ as follows: (i) $\beta_{1, 1} = \beta_H = (\sqrt{2} + 
1) \pi \sqrt{2\alpha^\prime}\: ; \; \tilde{\beta}_{1, 1} = \tilde{\beta}_H 
= 
(\sqrt{2} - 1)\pi \sqrt{2\alpha^\prime}\:$ , (ii) $\beta_{1, 3} = 
\tilde{\beta}_{3, 1} = \sqrt{3} \pi \sqrt{2\alpha^\prime}\: ; \; \beta_{3, 
1} 
= \tilde{\beta}_{1, 3} = 1/\sqrt{3} \cdot \pi 
\sqrt{2\alpha^\prime}\:$ .  In 
particular, $\beta_{1, 1}$ and $\tilde{\beta}_{1, 1}$ form the leading 
branch points of the square root type at $\beta_H$ and $\tilde{\beta}_H$, 
respectively.  Moverover, $\beta_H^{-1}$ $[\tilde{\beta}_H^{-1}]$ 
represents the lowest temperature singularity for the physical $\beta$ 
[dual $\tilde{\beta}$] channel.  Both $\beta_{1, 3}$ and $\beta_{3, 1}$ 
are ordinary points and consequently left out of consideration.  It is of 
practical importance to note that there exists no self-dual leading 
branch point at $\beta_0 = \tilde{\beta}_0 = \pi \sqrt{2\alpha^\prime}$ 
as well as any non-leading branch point on the physical sheet of 
the inverse temperature.  The present theoretical observation based upon 
the TFD free energy amplitude of the $D = 10$ heterotic thermal string 
yields a striking contrast to the previous argument by ourselves 
\cite{fujisaki4} for the $D = 26$ closed bosonic thermal string theory in 
the TFD framework.  We are now in the position to touch upon {\it 
$\grave{a}$ la} ref. \cite{fujisaki4} the global 
phase structure of the $D = 10$ heterotic thermal string ensemble.  
Analysis is performed $\grave{a}\: la$ ref. \cite{obrien}, ref. 
\cite{leblanc3} -- ref. \cite{deo} through the microcanonical ensemble 
paradigm outside the analyticity domain of the canonical ensemble.  There 
will then appear three phases in the sense of the thermal duality 
symmetry as follows \cite{fujisaki4}, \cite{obrien}, \cite{leblanc3}: (i) 
the $\beta$ channel canonical phase in the domain $(2 + \sqrt{2})\pi \sqrt
{\alpha^\prime} = \beta_H \leq \beta < \infty$, (ii) the dual 
$\tilde{\beta}$ channel canonical phase in the domain $0 < \beta \leq 
\tilde{\beta}_H = (2 - \sqrt{2})\pi \sqrt{\alpha^\prime}$ and 
(iii) the self-dual microcanonical phase in the domain $\tilde{\beta}_H < 
\beta 
< \beta_H$.  In sharp contrast to the global phase structure of the $D = 
26$ closed bosonic thermal string ensemble \cite{fujisaki4}, however, 
there 
will occur no effective splitting of the microcanonical region because of 
the absence of the self-dual branch point at $\beta_0 = \tilde{\beta}_0 
= \pi \sqrt{2\alpha^\prime}$ as well as any secondary singularity.  As a 
consequence of the self-duality of the microcanonical phase, therefore, 
it may be possible to speculate that the so-called maximum temperature of 
the $D = 10$ heterotic string excitation is asymptotically described at 
least at the one-loop level as $\tilde{\beta}_H^{-1}$ $[\beta_H^{-1}]$ in 
replacement of $\beta_0^{-1} = \tilde{\beta}_0^{-1}$ for the physical 
$\beta$ [dual $\tilde{\beta}$] channel.  

We have anyhow succeeded in shedding some light upon the global phase 
structure of the thermal string ensemble through the infrared 
behaviour of the one-loop free energy amplitude for the dimensionally 
regularized, $D = 10$ heterotic thermal string theory based upon the 
TFD algorithm.  In particular, the full use has been made of the thermal 
duality symmetry as well as the thermal stability of modular 
invariance not only for the canonical region but also for 
the microcanonical region.  It is hoped that we can illuminate the fruitful 
thermodynamical investigation of string excitations, {\it e.g.} the manifest 
materialization of the ``true'' maximum temperature for the thermal 
string ensemble in general within the new-fashioned duality framework of the
 $D$-brane paradigm \cite{mozo}.
\\

One of the authors (H. F.) is grateful to Prof. S. Saito for the 
hospitality of Tokyo Metropolitan University.    
%%%%%%%%
%%%%%%%%


\newpage
\begin{thebibliography}{99}
\bibitem{umezawa} See, for example, H. Umezawa, H. Matsumoto and M. 
Tachiki, {\it Thermo Field Dynamics and Condensed States} (North-Holland, 
Amsterdam, 1982).  For a recent publication, see, for example, P. A. 
Henning, {\it Phys. Rep.} {\bf 253} (1995) 235.
\bibitem{leblanc1} Y. Leblanc, {\it Phys. Rev. D} {\bf 36} (1987) 1780; 
{\it D} {\bf 37} (1988) 1547; {\it D} {\bf 39} (1989) 1139; 3731.
\bibitem{leblanc2} Y. Leblanc, M. Knecht and J. C. Wallet,  {\it Phys. 
Lett. B} {\bf 237} (1990)  357.
\bibitem{ahmed} E. Ahmed, {\it Int. J. Theor. Phys.} {\bf 26} (1988) 1135; 
{\it Phys. Rev. Lett.} {\bf 60} (1988) 684.
\bibitem{fujisaki1} H. Fujisaki, {\it Prog. Theor. Phys.} {\bf 81} (1989) 
473; {\bf 84} (1990) 191; {\bf 85} (1991) 1159; {\bf 86} (1991) 509; {\it 
Europhys. Lett.} {\bf 14} (1991) 737; {\bf 19} (1992) 73; {\bf 28} (1994), 
623; {\it Nuovo Cim.} {\bf 108}{\it A} (1995) 1079.
\bibitem{fujisaki2} H. Fujisaki, K. Nakagawa and I. Shirai, {\it Prog. 
Theor. Phys.} {\bf 81} (1989) 565; 570.
\bibitem{fujisaki3} H. Fujisaki and K. Nakagawa, {\it Prog. Theor. Phys.} 
{\bf 82} (1989) 236; 1017; {\bf 83} (1990) 18; {\it Europhys. Lett.} {\bf 
14} (1991) 639; {\bf 20} (1992) 677 ; {\bf 28} (1994) 1; {\it ibid.} 471.
\bibitem{fujisaki4} H. Fujisaki and K. Nakagawa, {\it Europhys. Lett.} 
{\bf 35} (1996) 493.
\bibitem{nakagawa} K. Nakagawa, {\it Prog. Theor. Phys.} {\bf 85} (1991) 
1317.
\bibitem{obrien} K. H. O'Brien and C. -I Tan, {\it Phys. Rev. D} {\bf 36} 
(1987) 1184.
\bibitem{atick} J. J. Atick and E. Witten, {\it Nucl. Phys. B} {\bf 310} 
(1988) 291.
\bibitem{alvarez} E. Alvarez and M. A. R. Osorio, {\it Nucl. Phys. B} 
{\bf 304} (1988) 327 ; {\it Phys. Rev. D} {\bf 40} (1989) 1150.  
\bibitem{osorio} M. A. R. Osorio, {\it Int. J. Mod. Phys. A} {\bf 7} (1992) 4275. 
\bibitem{green} See, for example, M. B. Green, J. H. Schwarz and E. 
Witten,  {\it Superstring Theory}, Vols. 1 and 2 (Cambridge Univ. Press, 
Cambridge, 1987).
\bibitem{leblanc3} Y. Leblanc, {\it Phys. Rev. D} {\bf 38} (1988) 3087.
\bibitem{brandenberger} R. Brandenberger and C. Vafa, {\it Nucl. 
Phys. B} {\bf 316} (1989) 391. 
\bibitem{deo} N. Deo, S. Jain and C. -I Tan, {\it Proceedings of the 
International Colloquium on Modern Quantum Field Theory}; ed. by S. Das 
{\it et al.} (World Scientific Pub. Co., Singapore, 1991), p. 112.
\bibitem{mozo} For a nearly related publication, see, for example, M. A. 
V$\acute{a}$zquez-Mozo, Princeton preprint IASSNS-HEP-96-73; 
hep-th/9607052 (1996).  
\end{thebibliography}
\end{document}